\documentclass[lettersize,journal]{IEEEtran}

\usepackage{cite}
\usepackage{amsmath,amssymb,amsfonts}
\usepackage{algorithmic}
\usepackage{graphicx}
\usepackage{algorithm,algorithmic}
\usepackage{hyperref}
\usepackage{booktabs}
\usepackage{multirow}
\hypersetup{hidelinks}
\usepackage{textcomp}
\usepackage[caption=false,font=footnotesize]{subfig}
\usepackage[percent]{overpic}

\def\BibTeX{{\rm B\kern-.05em{\sc i\kern-.025em b}\kern-.08em
    T\kern-.1667em\lower.7ex\hbox{E}\kern-.125emX}}
\begin{document}
\title{Interpretable Information-Decomposed Brain Graph Learning for fMRI-based Disease Diagnosis}
\author{Dengyi Zhao, Zhiheng Zhou, Zihan Wang, Guiying Yan and Xingqin Qi, \IEEEmembership{Member, IEEE}
\thanks{This work was supported in part by the National Natural Science Foundation of China No.12471330, the Shandong Provincial Natural Science Foundation No.ZR2025MS71 and the National Natural Science Foundation of China No.12231018. (Corresponding authors:Zhiheng Zhou, Xingqin Qi.)}
\thanks{Dengyi Zhao, Zhiheng Zhou and Xingqin Qi are with School of Mathematics and Statistics, Shandong University, Weihai 264209, China (e-mail: zhaodengyi@mail.sdu.edu.cn; zhouzhiheng@amss.ac.cn; wangyunping@mail.sdu.edu.cn; qixingqin@sdu.edu.cn).}
 \thanks{Zihan Wang is with Data Science Institute, Shandong University,
Jinan, 250100, China (e-mail: wangzihan@mail.sdu.edu.cn).}
\thanks{Guiying Yan is with Academy of Mathematics and Systems Science, University of Chinese Academy of Sciences, Beijing 100190, China (e-mail: yangy@amt.ac.cn).}
}

\maketitle
\begin{abstract}
Resting-state functional magnetic resonance imaging (rs-fMRI) has enabled non-invasive mapping of functional brain interactions for computer-aided diagnosis, yet most existing approaches reduce inter-regional relationships to correlation-based edge weights. Such representations capture co-fluctuation strength but obscure how information is shared across brain regions. Because brain disorders may disrupt not only connectivity strength but also the organization of redundancy, uniqueness and synergy, traditional functional connectivity may miss disease-relevant information structures. Here we introduce IID-GCN, an interpretable graph learning framework that decomposes rs-fMRI interactions into redundancy, uniqueness and synergy graphs using partial entropy decomposition. These information-specific graphs separately characterize shared, region-specific and jointly emergent components of brain activity. A multi-channel graph convolutional network then integrates the decomposed graphs through edge recalibration, cross-information interaction, ROI-attention readout and channel-attentive fusion. Across three datasets, IID-GCN consistently captures complementary diagnostic information beyond traditional  functional connectivity. The learned information profiles reveal disorder-specific patterns of altered redundancy, uniqueness and synergy, suggesting that brain diseases reshape functional information organization rather than merely changing connection strength. These results establish information-decomposed brain graphs as an interpretable representation for rs-fMRI-based diagnosis. Our code is available
at https://github.com/Zdy12/IID-GCN.
\end{abstract}

\begin{IEEEkeywords}
brain disease diagnosis; graph neural network; partial entropy decomposition; brain graph.
\end{IEEEkeywords}

\section{Introduction}
\label{sec:introduction}
\IEEEPARstart{B}{rain} disorders such as Alzheimer's disease, Parkinson's disease and autism spectrum disorder are increasingly understood as systems-level disturbances of distributed brain function rather than abnormalities confined to isolated regions \cite{fornito2015connectomics}. Resting-state functional magnetic resonance imaging (rs-fMRI) provides a non-invasive measure of spontaneous blood-oxygen-level-dependent activity and has become a central tool for mapping large-scale functional brain organization in health and disease \cite{fox2007spontaneous,van2010exploring}. In computer-aided diagnosis, a common strategy is to parcellate the brain into regions of interest, estimate inter-regional functional connectivity from regional time series, and use the resulting brain network as the basis for classification \cite{yin2022deep,teng2023brain}. This network view has enabled substantial progress, but it also raises a fundamental representational question: whether traditional  functional connectivity adequately captures the information organization underlying disease-related brain dysfunction.

Most rs-fMRI diagnostic studies construct brain networks using Pearson correlation or related pairwise association measures \cite{van2010exploring,hlinka2011functional,zhou2024novel,zhao2026hoi,zhao2026extracting,wang2026classification}. Although these measures are simple, robust and widely used, they reduce the relationship between two regions of interest (ROIs) to a single scalar reflecting linear co-fluctuation strength \cite{hlinka2011functional,hartman2011role}. Such a representation cannot distinguish qualitatively different modes of information sharing. Two regions may convey redundant information, one region may carry information unique from the other, or their joint activity may encode synergistic information that is not available from either region alone \cite{wibral2017partial}. These components may reflect distinct computational roles in distributed neural processing and may be differentially altered by disease \cite{luppi2024synergistic}. Recent advances in partial entropy decomposition provide a principled framework for separating these redundant, unique and synergistic information atoms, and accumulating evidence suggests that human brain activity contains rich synergistic structure that is not visible from traditional  pairwise connectivity alone \cite{varley2023partial}. Thus, collapsing heterogeneous information-sharing modes into a single correlation edge may obscure clinically meaningful alterations in functional brain organization.

Graph neural networks (GNNs) have become a leading approach for rs-fMRI-based diagnosis because they can learn directly from brain networks and identify disease-relevant patterns beyond handcrafted graph measures \cite{li2021braingnn,bessadok2022graph}. Existing models have improved this paradigm by incorporating ROI aware convolution, interpretable pooling, node grouping and subnetwork discovery \cite{yan2019groupinn,li2021braingnn}, and recent surveys have emphasized their growing role in brain connectivity analysis \cite{mohammadi2024graph,luo2024graph}. Yet these advances mainly address how to learn from a brain graph, while leaving a more basic question unresolved: what information should the graph encode? When the graph is built from Pearson functional connectivity, the input to the GNN is already a compressed representation in which redundancy, uniqueness and synergy are indistinguishable \cite{hlinka2011functional,wibral2017partial,varley2023partial}. The model may therefore improve classification accuracy without revealing which mode of information sharing is disrupted in disease. This representational bottleneck limits the mechanistic value of current graph-learning approaches and motivates a framework that explicitly constructs and jointly learns from redundancy, uniqueness and synergy graphs. How to achieve this within an end-to-end and interpretable diagnostic model remains an open problem.

To address this gap, we propose IID-GCN, an interpretable information-decomposed graph convolutional network for rs-fMRI-based brain disease diagnosis. Instead of constructing a single correlation-based functional connectivity graph, IID-GCN transforms regional rs-fMRI signals into binary brain-state sequences and applies  partial entropy decomposition to generate three subject-specific information graphs: redundancy, uniqueness, and synergy graphs. These complementary graphs are jointly modeled by a multi-channel GNN, where each channel is encoded by an independent graph convolutional branch. Residual edge recalibration adaptively refines information-specific connections, while node-wise cross-information interaction models the interplay among redundancy, uniqueness, and synergy at each ROI. ROI attention, soft node grouping, and channel-attentive fusion further identify disease-relevant regions, latent subnetworks, and the relative diagnostic contribution of each information component. Therefore, IID-GCN aims to improve classification while providing multi-level interpretability through channel weights, ROI-specific information preferences, attention maps, learned subnetworks.

We evaluate IID-GCN on three public neuroimaging datasets covering both neurodegenerative and neurodevelopmental disorders. This multi-dataset evaluation examines whether redundancy, uniqueness, and synergy provide disorder-specific information beyond traditional  functional connectivity. The experimental protocol includes diagnostic classification, disease- or development-stage analysis, component-level and module-level ablations, robustness and threshold sensitivity analyses, and interpretability studies. These experiments investigate whether information-decomposed brain graphs improve rs-fMRI-based diagnosis, whether different information components contribute differently across disorders, and whether their patterns vary along disease progression or developmental stage.




Our main contributions are:
\begin{itemize}
    \item We introduce an information-decomposed representation of rs-fMRI brain networks beyond traditional  functional connectivity.

    \item We propose IID-GCN, an interpretable multi-channel GCN that jointly learns information-specific brain graphs for brain disorder diagnosis.

    \item We evaluate IID-GCN on three public neuroimaging datasets showing its effectiveness, robustness, and interpretability across multiple brain disorders.
\end{itemize}

\section{RELATED WORKS}
\label{RELATED WORKS}
\subsection{Functional Network-Based Brain Disease Diagnosis}

Brain disorder diagnosis has increasingly relied on rs-fMRI to characterize spontaneous functional interactions among distributed ROIs. A standard pipeline constructs functional connectivity networks from regional BOLD time series and uses them for disease classification across disorders such as Alzheimer's disease, Parkinson's disease, autism spectrum disorder, and schizophrenia \cite{yin2022deep,teng2023brain}. Pearson correlation remains widely used because of its simplicity and interpretability, while partial correlation, mutual information, dynamic functional connectivity, and graph-theoretical descriptors have been explored as complementary measures.

Although recent studies confirm the diagnostic value of rs-fMRI connectivity in disease staging, dynamic network analysis, and connectome-based characterization \cite{alarjani2025brain,tessadori2025linking,karim2025functional}, most methods still encode each subject by a single connectivity matrix. Each edge is therefore reduced to one scalar association strength, which may obscure distinct modes of information sharing. Consequently, disease-related changes in redundancy, uniqueness, and synergy can be compressed into undifferentiated connectivity weights, motivating information-decomposed brain network representations.

\subsection{Graph Neural Networks for Brain Graph Learning}

GNNs have become widely used for brain connectivity analysis because they can jointly model regional attributes, network topology, and graph-level supervision. In rs-fMRI-based diagnosis, they have been applied to disease classification and disease-related ROI identification. Representative methods include BrainGNN with ROI-aware graph convolution and ROI selection pooling \cite{li2021braingnn}, as well as hypergraph, local-to-global, spatio-temporal, and contrastive graph learning approaches \cite{ji2022fchat,zhang2023local,wang2023ucgl}. Recent surveys further summarize the growing role of GNNs in functional brain network analysis and brain disease diagnosis \cite{tang2025gnnfmri,ali2025gnnad,sun2025admgcn}.

However, most existing methods mainly improve graph learning architectures while retaining Pearson correlation or related pairwise connectivity measures as input graphs. As a result, heterogeneous information-sharing components, including redundancy, uniqueness, and synergy, are collapsed into a single edge weight before learning. This representational bottleneck motivates IID-GCN, which constructs information-decomposed brain graphs and learns their complementary diagnostic contributions in an interpretable framework.

\begin{figure*}
    \centering
    \includegraphics[scale=0.55]{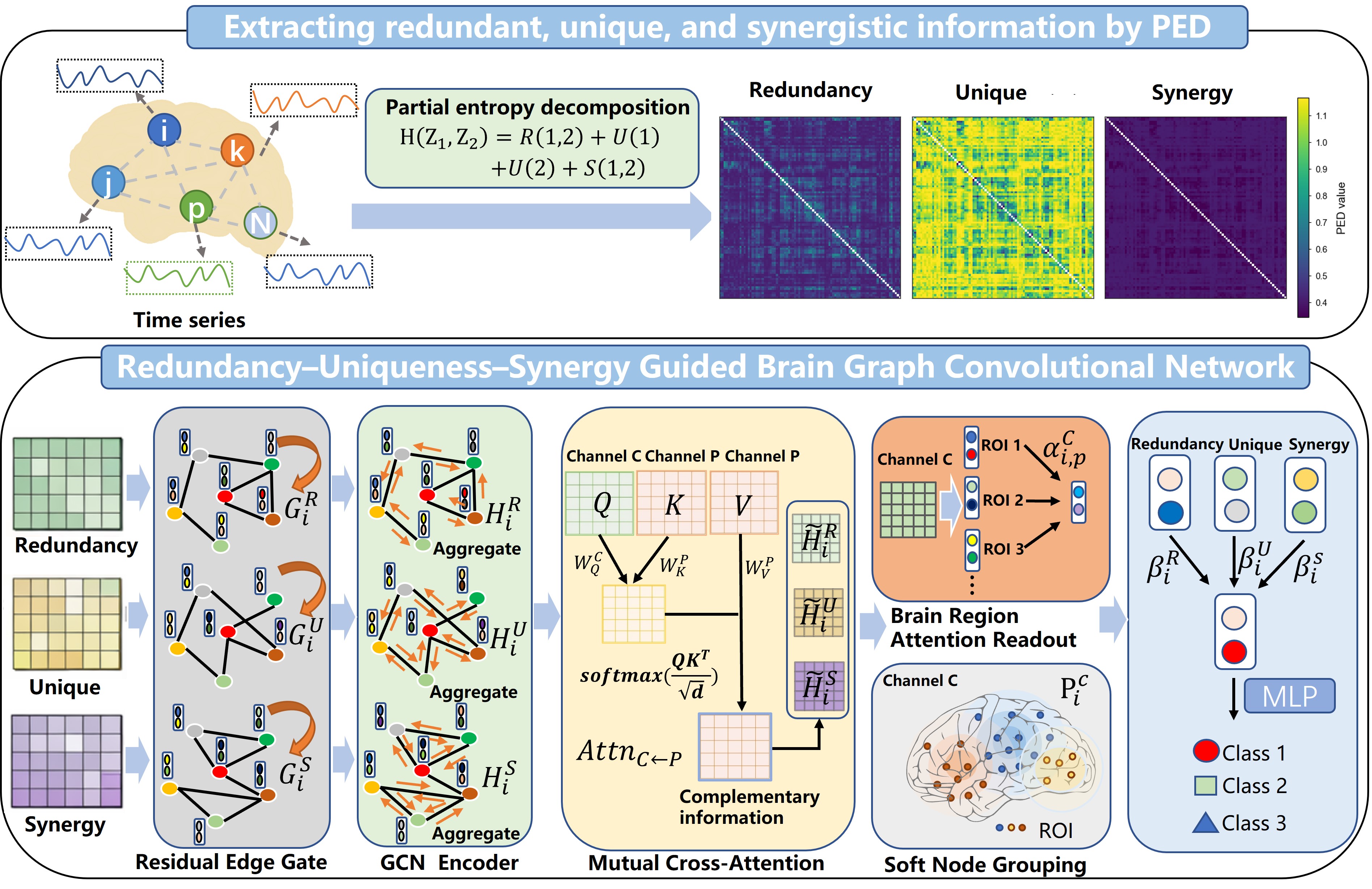}
    \caption{
    Overall framework of the proposed IID-GCN. The upper panel shows the construction of information-specific brain graphs from individual rs-fMRI time series. For each subject, regional rs-fMRI signals are binarized into brain-state sequences, and partial entropy decomposition (PED) is applied to each ROI pair to obtain redundancy, uniqueness, and synergy components, forming three subject-specific information brain networks. The lower panel illustrates the IID-GCN architecture. The three information networks are first refined by a residual edge gate and encoded by channel-specific GCN encoders. A node-wise cross-information interaction module models ROI-level dependencies among redundancy, uniqueness, and synergy representations. ROI attention readout and soft node grouping pooling then extract region-level and subnetwork-level graph representations, which are adaptively fused by channel attention for disease classification.
}
    \label{fig1}
\end{figure*}
\section{PROPOSED METHOD}
\label{PROPOSED METHOD}
\subsection{Overview}
Functional interactions among ROIs may involve distinct modes of information sharing. Motivated by this perspective, we propose IID-GCN, an interpretable information-decomposed brain graph convolutional network for rs-fMRI-based brain disease diagnosis. The overall architecture of IID-GCN is illustrated in Fig.~\ref{fig1}. In the following sections, we introduce each component of the proposed model in detail.

\subsection{Construction of Information-Specific Brain Graphs}
To obtain a more informative graph representation, we decompose pairwise regional dependencies into redundancy, uniqueness, and synergy. This construction has two advantages. First, it separates shared, region-specific, and jointly emergent information components that are collapsed in correlation-based connectivity. Second, it provides three information-specific brain graphs that can be jointly modeled by a multi-channel GNN, allowing the model to learn not only disease-discriminative patterns but also the relative contribution of different information components. Thus, instead of treating each subject as a single homogeneous functional graph, we represent each subject by a three-channel information-decomposed brain graph.

Given a labeled dataset
\(
\mathcal{D}=\{(\mathbf{X}_i,y_i)\}_{i=1}^{N},
\)
where \(N\) denotes the number of subjects, each subject is represented by rs-fMRI signals $\mathbf{X}_i$, together with a diagnostic label \(y_i\in\mathcal{Y}\). For each subject \(i\), let
\(
\mathbf{X}_i=\bigl[\mathbf{x}_i^1,\ldots,\mathbf{x}_i^M\bigr]\in\mathbb{R}^{M\times T}
\), where \(M\) is the number of ROIs, \(T\) is the number of time points, and \({x}_i^r(t)\) denotes the BOLD signal of region \(r\) at time point \(t\). Each regional time series is first standardized as
\begin{equation}
    \widetilde{x}_i^r(t)
    =
    \frac{x_i^r(t)-\mu_i^r}{\sigma_i^r+\epsilon},
\end{equation}
where \(\mu_i^r\) and \(\sigma_i^r\) are the mean and standard deviation of the time series of region \(r\), and \(\epsilon\) is a small constant for numerical stability. The standardized signal is then converted into a binary brain-state sequence by assigning values greater than zero to one and values less than or equal to zero to zero:
\begin{equation}
    b_i^r(t)
    =
    \mathbb{I}\left(\widetilde{x}_i^r(t)>0\right),
\end{equation}
where \(\mathbb{I}(\cdot)\) denotes the indicator function, which equals one if the condition inside the parentheses is true and zero otherwise.
For each subject \(i\) and ROI \(r\), we denote the resulting sequence as
\(
    B_i^r=\{b_i^r(t)\}_{t=1}^{T}.
\) This step provides a discrete representation suitable for entropy-based information decomposition.

We then use partial entropy decomposition (PED) to construct information-specific brain graphs \cite{ince2017partial}. For two discrete variables \(Z_1\) and \(Z_2\), PED decomposes their joint entropy into four interpretable information atoms:
\begin{equation}
    H(Z_1,Z_2)
    =
    R(1,2)
    +
    U(1)
    +
    U(2)
    +
    S(1,2).
\end{equation}
Here, \(R(1,2)\) denotes the redundant entropy shared by \(Z_1\) and \(Z_2\); \(U(1)\) and \(U(2)\) denote the unique entropy contributed by \(Z_1\) and \(Z_2\), respectively; and \(S(1,2)\) denotes the synergistic entropy that can be resolved from their joint state. The detailed computation of PED is described in \cite{varley2023partial}.

In this study, we apply PED to each pair of binary state sequences to quantify redundancy, uniqueness, and synergy between ROIs. For subject \(i\) and each pair of distinct regions \((p,q)\), we treat the \(
    B_i^p=\{b_i^p(t)\}_{t=1}^{T},\)
and
\(
    B_i^q=\{b_i^q(t)\}_{t=1}^{T}
\)
as two discrete variables. Their joint distribution is estimated from the co-occurrence frequencies of binary states across time points. Applying the PED yields a redundant component \(R_i(p,q)\), two directional unique components \(U_i(p)\) and \(U_i(q)\), and a synergistic component \(S_i(p,q)\). To obtain a symmetric uniqueness score for graph construction, we define
\(
    U_i(p,q)=\frac{1}{2}\left(U_i(p)+U_i(q)\right).
\)
Repeating this procedure for all unequal pairs of ROIs and setting the diagonal entries to zero yields three subject-specific information matrices, denoted by
\(
\mathbf{F}_i^R, \mathbf{F}_i^U
\)
and
\(
\mathbf{F}_i^S
\).
The value of each matrix entry is given by
\begin{equation}
\mathbf{F}_{i}^C(p,q)=
\begin{cases}
C_i(p,q),
& p\neq q,
\\[0.2cm]
0,
&
\text{otherwise},
\end{cases}
\quad
C\in\{R,U,S\}.
\end{equation}

To reduce noisy weak connections and improve graph sparsity, the final PED-based
brain graphs were constructed from
\(
\mathbf{F}_i^R,
\mathbf{F}_i^U
\)
and
\(
\mathbf{F}_i^S
\)
by retaining only the top-\(k\) strongest connections for each ROI in each information channel. The sparse adjacency matrices
\(
\mathbf{A}_i^R,
\mathbf{A}_i^U
\)
and
\(
\mathbf{A}_i^S
\)
were obtained as
\begin{equation}
\mathbf{A}_i^C(p,q)
=
\begin{cases}
\mathbf{F}_i^C(p,q),
&
q\in \mathcal{N}_{k}^{C}(p) \ \text{or} \ p\in \mathcal{N}_{k}^{C}(q),
\\[0.1cm]
0,
&
\text{otherwise},
\end{cases}
\end{equation}
where 
\(
\mathcal{N}_k^C(p),\mathcal{N}_{k}^{C}(q)
\)
denotes the set of top-\(k\) strongest neighboring ROIs of region \(p,q\) under channel \(C\).
The redundancy, uniqueness, and synergy graphs $\mathbf{A}_i^R$, $\mathbf{A}_i^U$, and $\mathbf{A}_i^S$ respectively characterize shared, region-specific, and jointly emergent information between ROIs, providing complementary descriptions of functional information organization beyond traditional  correlation-based connectivity.

\subsection{Residual Edge Gate}
The PED-derived information graphs provide subject-specific estimates of redundancy, uniqueness, and synergy between ROIs. However, not all information-specific connections are equally informative for diagnosis. Some edges may capture disease-relevant alterations in functional information organization, whereas others may reflect noise, weak associations, or non-discriminative variability. To adaptively identify diagnostically relevant connections while preserving the original PED-derived graph structure, we introduce a residual edge gate. This module learns continuous edge-wise modulation scores for each information channel and uses them to selectively up-weight or down-weight graph connections during training.

For subject \(i\) and information channel \(C\in\{R,U,S\}\), the raw edge gate is defined as
\begin{equation}
    \mathbf{G}_i^C
    =
    \sigma
    \left(
    s^C\mathbf{F}_i^C+\mathbf{M}^C
    \right),
\end{equation}
where \(s^C\) is a learnable scaling parameter, \(\mathbf{M}^C\in\mathbb{R}^{M\times M}\) is a learnable symmetric modulation matrix with zero diagonal entries, and \(\sigma(\cdot)\) denotes the sigmoid function.

The recalibrated adjacency matrix is constructed as:
\begin{equation}
    \widetilde{\mathbf{A}}_i^C
    =
    \mathbf{A}_i^C +\mathbf{A}_i^C
    \odot   \eta^C
    \left( 2\mathbf{G}_i^C-\mathbf{J}_{M\times M}
    \right),
\end{equation}
where \(\odot\) denotes element-wise multiplication, $\mathbf{J}_{M\times M}$ is a ${M\times M}$ all-ones matrix with zero diagonal entries, and \(
    \eta^C=\eta_{\max}\tanh(\rho^C)
\)
is a learnable channel-specific residual modulation coefficient. Here, \(\eta_{\max}>0\) controls the maximum modulation strength, and \(\rho^C\) is initialized to zero, which gives \(
    \widetilde{\mathbf{A}}_i^C=\mathbf{A}_i^C.
\) This initialization ensures that the model starts from the original PED-derived information graph and only gradually learns task-driven deviations from it. The bounded coefficient \(\eta^C\) prevents overly aggressive changes to the graph topology and improves training stability. 


\subsection{Information-Specific Graph Convolution Layer}
After edge-level recalibration, each information-specific brain graph is encoded by an independent graph convolutional branch. Graph convolution updates each ROI representation by aggregating information from its graph-defined neighbors, thereby capturing local and higher-order connectivity patterns in the brain network. Moreover, using separate branches for redundancy, uniqueness, and synergy allows the model to learn channel-specific functional organization without prematurely mixing heterogeneous information components.

For subject \(i\) and information channel \(C\in\{R,U,S\}\), the initial node embedding is obtained by
\(
    \mathbf{H}_i^{C,0}
    =
    \phi_{\mathrm{in}}^C
    \left(
    \mathbf{F}_i^C
    \right),
\)
where \(\phi_{\mathrm{in}}^C(\cdot)\) is a channel-specific input projection layer. 
Before graph convolution, symmetric normalization of the recalibrated adjacency matrix is applied \(
    \widehat{\mathbf{A}}_i^C
    =
    (\widetilde{\mathbf{D}}_i^C)^{-\frac{1}{2}}
    \left(
    \widetilde{\mathbf{A}}_i^C+\mathbf{I}
    \right)
    (\widetilde{\mathbf{D}}_i^C)^{-\frac{1}{2}},
\)
where \(\mathbf{I}\) is the identity matrix and \(\widetilde{\mathbf{D}}_i^C\) is the degree matrix of
\(\widetilde{\mathbf{A}}_i^C+\mathbf{I}\). The graph convolution at layer \(\ell\) is then defined as
\begin{equation}
    \mathbf{H}_i^{C,\ell+1}
    =
    \operatorname{ReLU}
    \left(
    \widehat{\mathbf{A}}_i^C
    \mathbf{H}_i^{C,\ell}
    \mathbf{W}_{\ell}^{C}
    \right),
\end{equation}
where \(\mathbf{W}_{\ell}^{C}\) is a learnable weight matrix. Two graph convolutional layers are used in each information channel.

The final node representations of the three channels are denoted by
\(\mathbf{H}_i^{R},\mathbf{H}_i^{U},\) and \(\mathbf{H}_i^{S},\) which encode channel-specific patterns of functional information organization.

\subsection{Mutual Cross-Attention-Based Information Fusion}
To model the complementary relationships among redundancy, uniqueness, and synergy, we introduce a mutual cross-attention-based information fusion module. The key idea is to fix one information channel as the target branch and use the other two channels as complementary source branches. In this way, each channel can preserve its original information while selectively absorbing complementary node-level information from the remaining channels.

For each target channel \(C\), we use its node representation \(\mathbf{H}_i^C\) as the query and use another channel \(P\) as the source of complementary information. The cross-channel attention is defined as
\begin{equation}
    \operatorname{Attn}_{C\leftarrow {P}}
    =
    \operatorname{softmax}
    \left(
    \dfrac{
    \left(\mathbf{H}_i^C\mathbf{W}_Q^C\right)
    \left(\mathbf{H}_i^{P}\mathbf{W}_K^{P}\right)^{\top}
    }{
    \sqrt{d}
    }
    \right)
    \left(\mathbf{H}_i^{P}\mathbf{W}_V^{P}\right),
\end{equation}
where \(\mathbf{W}_Q^C\), \(\mathbf{W}_K^P\), and \(\mathbf{W}_V^P\) are learnable projection matrices, and \(d\) is the hidden feature dimension. 
In this formulation, the attention weights quantify how ROIs in the target information channel attend to ROIs in a complementary source channel.

The refined representations of the three information channels are obtained such that each channel preserves its original representation while incorporating cross-attended information from the other two channels:
\begin{align}
    \widetilde{\mathbf{H}}_i^R
    &=
    \mathbf{H}_i^R
    +
    \operatorname{Attn}_{R\leftarrow U}
    +
    \operatorname{Attn}_{R\leftarrow S}, \\
    \widetilde{\mathbf{H}}_i^U
    &=
    \mathbf{H}_i^U
    +
    \operatorname{Attn}_{U\leftarrow R}
    +
    \operatorname{Attn}_{U\leftarrow S}, \\
    \widetilde{\mathbf{H}}_i^S
    &=
    \mathbf{H}_i^S
    +
    \operatorname{Attn}_{S\leftarrow R}
    +
    \operatorname{Attn}_{S\leftarrow U}.
\end{align}

\subsection{Information-Guided Graph Readout and Prediction}
After cross-information interaction, IID-GCN aggregates the refined node representations into graph-level representations for disease prediction. For each information channel \(C\in\{R,U,S\}\), the refined node representation of subject \(i\) is denoted by
\(
    \widetilde{\mathbf{H}}_i^C
    =
    \bigl[
    \widetilde{\mathbf{h}}_{i,1}^C,
    \ldots,
    \widetilde{\mathbf{h}}_{i,M}^C
    \bigr]
    \in\mathbb{R}^{M\times d},
\)
where \(d\) is the hidden feature dimension. To capture complementary graph-level information, we combine global pooling, ROI-attention readout, and soft subnetwork-level grouping.

First, global mean pooling and max pooling are applied to summarize the overall distribution of ROIs representations:
\begin{equation}
    \mathbf{g}_{i,\mathrm{mean}}^C
    =
    \operatorname{Mean}
    \left(
    \widetilde{\mathbf{H}}_i^C
    \right),
    \quad
    \mathbf{g}_{i,\mathrm{max}}^C
    =
    \operatorname{Max}
    \left(
    \widetilde{\mathbf{H}}_i^C
    \right).
\end{equation}

Second, to identify disease-relevant ROIs, we introduce an ROI-attention readout. For region \(p\), the attention score is computed as
\begin{equation}
    e_{i,p}^C
    =
    \mathbf{v}_e^\top
    \tanh
    \left(
        \mathbf{W}_e^C \widetilde{\mathbf{h}}_{i,p}^C
    \right),
\end{equation}
where \(\mathbf{W}_e^C\) and \(\mathbf{v}_e\) are learnable parameters. The normalized ROI attention weight and the ROI attentive graph representation are given by
\begin{equation}
    \alpha_{i,p}^C
    =
    \frac{
        \exp(e_{i,p}^C)
    }{
        \sum_{q=1}^{M}
        \exp(e_{i,q}^C)
    },\quad\mathbf{g}_{i,\mathrm{region}}^C
    =
    \sum_{p=1}^{M}
    \alpha_{i,p}^C
    \widetilde{\mathbf{h}}_{i,p}^C.
\end{equation}
The learned weights \(\alpha_{i,p}^C\) provide an ROI level explanation of which ROIs are emphasized in each information channel.

Third, to capture subnetwork-level organization, we use soft node grouping. The soft assignment matrix is computed by
\begin{equation}
    \mathbf{P}_i^C
    =
    \operatorname{softmax}
    \left(
        \operatorname{MLP}^{C}
        \left(
        \widetilde{\mathbf{H}}_i^C
        \right)
    \right),
    \quad
    \mathbf{P}_i^C\in\mathbb{R}^{M\times K},
\end{equation}
where \(K\) is the number of latent brain groups. The group-level representation $\mathbf{g}_{i,\mathrm{group}}^C$ is obtained as
\begin{equation}
    \mathbf{g}_{i,\mathrm{group}}^C
    =
    \operatorname{vec}((\mathbf{P}_i^C)^\top
    \widetilde{\mathbf{H}}_i^C)
    \in\mathbb{R}^{Kd},
\end{equation}
where $\mathrm{vec}(\cdot)$ denotes the matrix vectorization operation.
This mechanism allows the model to summarize latent subnetworks rather than relying only on ROI level features.

The channel-specific graph representation is obtained by concatenating the four readout components:
\begin{equation}
    \mathbf{g}_i^C
    =
    \operatorname{Concat}
    \left(
        \mathbf{g}_{i,\mathrm{mean}}^C,
        \mathbf{g}_{i,\mathrm{max}}^C,
        \mathbf{g}_{i,\mathrm{region}}^C,
        \mathbf{g}_{i,\mathrm{group}}^C
    \right).
\end{equation}
Thus, each information channel is represented by a graph-level embedding that combines multi-level information.

To adaptively integrate redundancy, uniqueness, and synergy, we further introduce channel attention. The attention score of channel \(C\) is computed as
\begin{equation}
    f_i^C
    =
    \mathbf{q}^{\top}
    \tanh
    \left(
        \mathbf{W}^C_{f}
        \mathbf{g}_i^C
    \right).
\end{equation}
The normalized channel attention weight and the channel-attentive representation are defined as
\begin{equation}
    \beta_i^C
    =
    \frac{
        \exp(f_i^C)
    }{
        \sum\limits_{C'\in\{R,U,S\}}
        \exp(f_i^{C'})
    },\, \mathbf{g}^{all}_{i}
    =
    \sum_{C\in\{R,U,S\}}
    \beta_i^C
    \mathbf{g}_i^C.
\end{equation}
The weights \(\beta_i^C\) quantify the relative contribution of redundancy, uniqueness, and synergy to the final prediction.

The fused graph-level representation $\mathbf{g}^{all}_{i}$ is fed into a classification layer to obtain the predicted diagnostic probability $\widehat{\mathbf{p}}_i$. This readout and fusion design enables IID-GCN to integrate complementary information from ROI-level patterns, latent subnetworks, and information channels.

\subsection{Training Objective}

IID-GCN is trained in a supervised manner using a  cross-entropy loss. For a training set with \(N\) subjects, the classification loss is defined as
\begin{equation}
   \mathcal{L} =
    -
    \sum_{i=1}^{N}
    \sum_{m=1}^{K_c}
    \mathbb{I}(y_i=m)
    \log \widehat{\mathbf{p}}_{i,m},
\end{equation}
where \(K_c\) denotes the number of diagnostic classes, and \(\widehat{\mathbf{p}}_{i,m}\) is the predicted probability that subject \(i\) belongs to class \(m\).


\begin{table}[h]
\centering
\caption{Class distribution of brain network datasets.}
\vspace{2pt}
\label{tab:class_distribution}
\scalebox{0.74}{
\begin{tabular}{@{}lllc@{}}
\toprule
Dataset & Class & \# Subjects & Disease Type \\
\midrule

\multirow{3}{*}{\parbox{3cm}{ ADNI}} & AD & 90 & \multirow{3}{*}{\parbox{5cm}{\centering Alzheimer's Disease}} \\
     & MCI & 76 &  \\
     & CN & 96 &  \\
\midrule
{\parbox{3cm}{ PPMI}} & PD & 53 & \multirow{2}{*}{\parbox{5cm}{\centering Parkinson's Disease}} \\
     &  prodromal & 53 &  \\ 
     \midrule
\multirow{2}{*}{\parbox{3cm}{ ABIDE }} &ASD & 488 & \multirow{2}{*}{\parbox{5cm}{\centering Autism Spectrum Disorder }}  \\
      & CN & 537 &  \\
 \bottomrule

\end{tabular}
}
\end{table}

\begin{table*}[ht]
  \centering
  \caption{Performance comparison with  baselines on the ADNI, PPMI, and ABIDE datasets (\%). The best results are marked in bold and the second-best results are underlined.}
  \label{tab:performance_comparison}
  \scalebox{0.78}{
  \begin{tabular}{@{}lcccccccccccc@{}}
    \toprule
    \multirow{2}{*}{Method} 
    & \multicolumn{4}{c}{ADNI} 
    & \multicolumn{4}{c}{PPMI} 
    & \multicolumn{4}{c}{ABIDE} \\
    \cmidrule(lr){2-5} \cmidrule(lr){6-9} \cmidrule(lr){10-13}
    & Accuracy & Precision & Recall & F1-score
    & Accuracy & Precision & Recall & F1-score
    & Accuracy & Precision & Recall & F1-score \\
    \midrule
    MLP 
    &  59.3$\pm$8.8 & 57.7$\pm$8.4 & 58.4$\pm$7.4 & 58.1$\pm$5.8 & 61.3$\pm$5.3 & \underline{65.3$\pm$11.0} & 54.5$\pm$11.4 & 58.1$\pm$6.4  & 58.3$\pm$6.3 & 60.3$\pm$5.0 & 59.6$\pm$5.4 & 58.7$\pm$6.4\\
    SVM & 64.4$\pm$5.5 & 65.5$\pm$5.8 & 64.4$\pm$5.5 & 63.8$\pm$5.9 & \underline{64.2$\pm$8.3} & 58.4$\pm$10.4 & 58.4$\pm$10.4 & 61.8$\pm$8.7& 61.4$\pm$5.5 & 63.8$\pm$5.6 & 61.4$\pm$5.1 & 62.5$\pm$5.0 \\
    GCN& 60.9$\pm$11.8 & 61.6$\pm$12.5 & 61.1$\pm$11.9 & 60.3$\pm$12.1& 57.5$\pm$6.8 & 58.7$\pm$10.1 & 62.4$\pm$5.1 & 59.6$\pm$3.7 & 60.4$\pm$4.5 & 61.8$\pm$4.3 & \underline{64.4$\pm$7.5} & \underline{62.9$\pm$4.7}\\
     GraphSAGE & 64.5$\pm$8.9 & 65.6$\pm$9.1 & 64.5$\pm$8.9 & 64.0$\pm$9.0& 60.4$\pm$5.6 & 61.0$\pm$8.1 & 60.6$\pm$10.7 & 60.2$\pm$6.5 & 61.2$\pm$3.6 & \underline{64.0$\pm$4.1} & 59.3$\pm$5.6 & 61.5$\pm$5.5\\
      GAT & 60.0$\pm$10.1 & 61.8$\pm$10.6 & 60.0$\pm$10.1 & 59.2$\pm$10.3 & 61.3$\pm$6.8 & 61.3$\pm$7.3 & 64.4$\pm$8.1 & 60.4$\pm$6.3 & 59.3$\pm$3.8 & 57.9$\pm$5.3 & 61.1$\pm$3.5 & 59.5$\pm$4.8\\
      GroupINN & 57.3$\pm$8.8 & 61.0$\pm$9.6 & 57.4$\pm$9.8 & 56.9$\pm$9.1 & 55.6$\pm$6.1 & 55.7$\pm$6.2 & 55.6$\pm$6.0 & 55.4$\pm$6.2& 57.1$\pm$4.4 & 58.4$\pm$5.0 & 57.1$\pm$4.2 & 55.7$\pm$4.3 \\
     FBNETGEN & 66.4$\pm$7.0 & 67.7$\pm$7.5 & 67.4$\pm$7.1 & 60.8$\pm$5.9 & 59.4$\pm$8.4 & 60.6$\pm$10.9 & 62.4$\pm$5.1 & 60.8$\pm$5.9& 58.0$\pm$4.2 & 59.8$\pm$4.1 & 61.1$\pm$8.6 & 60.1$\pm$5.2 \\
    BPI-GNN & 51.4$\pm$8.5 & 49.2$\pm$14.6 & 52.1$\pm$8.8 & 48.0$\pm$11.1 & 52.3$\pm$14.7 & 50.5$\pm$18.9 & 49.4$\pm$15.1 & 47.5$\pm$14.9& 52.5$\pm$4.7 & 51.7$\pm$6.9 & 54.6$\pm$5.5 & 53.5$\pm$4.5 \\
    BrainGNN & 68.3$\pm$5.4 & 69.6$\pm$6.8 & 66.9$\pm$6.7 & 68.4$\pm$6.3 & 60.3$\pm$11.3 & 59.8$\pm$11.8 & 60.5$\pm$11.6 & 60.3$\pm$11.3 & 59.8$\pm$5.5 & 59.5$\pm$5.5 & 59.9$\pm$5.7 & 59.8$\pm$5.5 \\
      ATPGCN & \underline{70.8$\pm$6.7} & \underline{72.6$\pm$7.3} & \underline{70.6$\pm$6.3} & 70.8$\pm$6.7 & 61.2$\pm$3.5 & 62.1$\pm$3.4 & \underline{64.5$\pm$3.3} & \underline{62.9$\pm$3.3}& 62.2$\pm$4.5 & 62.4$\pm$5.3 & 60.6$\pm$5.3 & 61.8$\pm$4.6 \\
     Ada-MST & 68.8$\pm$6.3 & 70.5$\pm$7.1 & 68.6$\pm$5.7 & \underline{71.4$\pm$5.7} & \underline{62.7$\pm$5.6} & 62.0$\pm$3.1 & 61.3$\pm$5.4 &61.4$\pm$5.5& \underline{62.7$\pm$5.6} & 62.0$\pm$3.1 & 61.3$\pm$5.4 &61.4$\pm$5.5 \\
    \midrule
    IID-GCN & \textbf{73.9$\pm$5.8} & \textbf{75.2$\pm$5.1} & \textbf{73.4$\pm$5.5} & \textbf{73.4$\pm$5.2} & \textbf{70.7$\pm$12.2} & \textbf{71.6$\pm$12.3} & \textbf{70.7$\pm$12.2} & \textbf{70.3$\pm$12.5}& \textbf{65.1$\pm$3.6} & \textbf{65.0$\pm$3.2} & \textbf{65.0$\pm$3.6} & \textbf{64.9$\pm$3.5}\\
    \bottomrule
  \end{tabular}
  }
\end{table*}

\begin{table*}[ht]
  \centering
  \caption{Ablation study of different information-theoretic features with SVM on the ADNI, PPMI, and ABIDE datasets (\%). The best results are marked in bold and the second-best results are underlined.}
  \label{tab:feature_ablation_svm}
  \scalebox{0.78}{
  \begin{tabular}{@{}lcccccccccccc@{}}
    \toprule
    \multirow{2}{*}{Method} 
    & \multicolumn{4}{c}{ADNI} 
    & \multicolumn{4}{c}{PPMI} 
    & \multicolumn{4}{c}{ABIDE} \\
    \cmidrule(lr){2-5} \cmidrule(lr){6-9} \cmidrule(lr){10-13}
    & Accuracy & Precision & Recall & F1-score
    & Accuracy & Precision & Recall & F1-score
    & Accuracy & Precision & Recall & F1-score \\
    \midrule
    Pearson+SVM 
    & 64.4$\pm$5.5 & 65.5$\pm$5.8 & 64.4$\pm$5.5 & 63.8$\pm$5.9
    & 64.2$\pm$8.3 & 58.4$\pm$10.4 & 58.4$\pm$10.4 & 61.8$\pm$8.7
    & \underline{61.4$\pm$5.5} & \textbf{63.8$\pm$5.6} & 61.4$\pm$5.1 & \underline{62.5$\pm$5.0} \\

    Redundancy+SVM
    & 65.4$\pm$8.4 & 67.9$\pm$6.9 & 65.4$\pm$8.4 & 64.7$\pm$9.0
    & \underline{65.1$\pm$10.5} & \underline{63.2$\pm$11.5} & \underline{63.9$\pm$11.7} & \underline{64.8$\pm$12.5}
    & 60.7$\pm$3.1 & \underline{62.8$\pm$3.4} & 61.4$\pm$5.4 & 62.0$\pm$3.7 \\

    Unique+SVM
    & \underline{67.4$\pm$5.0} & \underline{68.5$\pm$4.7} & \underline{67.4$\pm$5.0} & \underline{67.0$\pm$5.0}
    & 60.0$\pm$7.3 & 61.9$\pm$9.3 & 60.0$\pm$7.3 & 58.3$\pm$10.0
    & 60.6$\pm$3.0 & 62.7$\pm$3.2 & 60.6$\pm$3.0 & 61.4$\pm$5.4 \\

    Synergy+SVM
    & 63.8$\pm$6.3 & 65.8$\pm$6.3 & 63.8$\pm$6.3 & 63.2$\pm$6.5
    & 63.4$\pm$8.5 & 61.2$\pm$9.7 & 61.5$\pm$10.6 & 63.3$\pm$11.4
    & 60.2$\pm$2.7 & 62.4$\pm$2.9 & \underline{62.4$\pm$2.9} & 61.4$\pm$3.3 \\

    All+SVM
    & \textbf{68.9$\pm$5.8} & \textbf{70.2$\pm$5.1} & \textbf{68.9$\pm$5.8} & \textbf{68.4$\pm$5.7}
    & \textbf{65.7$\pm$8.4} & \textbf{64.5$\pm$9.2} & \textbf{64.7$\pm$12.3} & \textbf{67.3$\pm$12.0}
    & \textbf{61.7$\pm$4.6} & 62.1$\pm$5.0 & \textbf{65.5$\pm$4.2} & \textbf{63.6$\pm$3.7} \\
    \bottomrule
  \end{tabular}
  }
\end{table*}
\section{EXPERIMENTS}
\label{EXPERIMENTS}

\subsection{Datasets and Preprocessing}

We evaluated the proposed method on three publicly available datasets covering neurodegenerative and neurodevelopmental disorders. The class distributions of the three datasets are summarized in Table~\ref{tab:class_distribution}. The Alzheimer's Disease Neuroimaging Initiative (ADNI) dataset \cite{dadi2019benchmarking} includes Alzheimer's disease (AD), mild cognitive impairment (MCI), and cognitively normal control (CN) subjects. The Parkinson's Progression Markers Initiative (PPMI) dataset \cite{xu2023data} includes Parkinson's disease (PD) and prodromal subjects. The Autism Brain Imaging Data Exchange (ABIDE) dataset \cite{craddock2013neuro} includes autism spectrum disorder (ASD) and normal control (CN) subjects from multiple international sites. All data were preprocessed following the fMRIPrep pipeline \cite{esteban2019fmriprep}. The preprocessed images were then parcellated using the Automated Anatomical Labeling (AAL) atlas \cite{tzourio2002automated}, resulting in 116 regions of interest. We further retained the 90 cerebrum ROIs for subsequent brain network construction and analysis.

\subsection{Baselines and Evaluation Metrics}

To evaluate IID-GCN, we compared it with representative baselines on ADNI, PPMI, and ABIDE, including traditional  classifiers, general-purpose GNNs, and brain-network-oriented deep models. Specifically, MLP and SVM \cite{pedregosa2011scikit} were used as vector-based classifiers with flattened brain network features. General GNN baselines included GCN \cite{kipf2016semi}, GraphSAGE \cite{hamilton2017inductive}, and GAT \cite{velivckovic2017graph}. Brain-network-specific baselines included GroupINN \cite{yan2019groupinn}, FBNETGEN \cite{kan2022fbnetgen}, BPI-GNN \cite{zheng2024bpi}, BrainGNN \cite{li2021braingnn}, ATPGCN \cite{bian2023adversarially}, and Ada-MST \cite{zeng2025adaptive}. All methods were evaluated under the same cross-validation protocol. Performance was measured by accuracy, precision, recall, and F1-score, with weighted averages used for multi-class tasks.

\subsection{Implementation Details }
For dataset partitioning, we adopted a 10-fold cross-validation protocol for all three datasets. In each run, the data were split into training, validation, and test subsets with an 8:1:1 ratio. Model selection was performed on the validation set, and the model achieving the highest validation F1-score was retained for testing. The results are reported as mean \(\pm\) standard deviation across folds. All models were trained using the Adam optimizer \cite{kingma2014adam} with a learning rate of \(10^{-3}\). 
The batch size was set to 16 for ADNI and ABIDE, and to 4 for PPMI due to its smaller sample size. The maximum number of training epochs was set to 200. 

\subsection{Performance Comparison}

Table~\ref{tab:performance_comparison} summarizes the classification results on all datasets. Overall, IID-GCN achieves the best mean performance across all three datasets. It obtains accuracies of \(73.9\%\), \(70.7\%\), and \(65.1\%\) on ADNI, PPMI, and ABIDE, respectively, outperforming the strongest baselines by \(3.1\%\), \(6.5\%\), and \(2.4\%\). Similar improvements are observed in precision, recall, and F1-score, indicating stable gains across different diagnostic tasks.

Compared with traditional  machine learning classifiers and generic GNNs, IID-GCN consistently performs better, suggesting that standard vector-based features or vanilla message passing are insufficient to fully capture disease-relevant functional patterns. IID-GCN also outperforms brain-network-oriented deep models, achieving the highest F1-scores on all datasets. These results demonstrate that information-decomposed brain graphs provide complementary diagnostic signals by separately modeling redundancy, uniqueness, and synergy, leading to a more expressive representation for rs-fMRI-based disease classification.

\subsection{Ablation Study}
\subsubsection{Effectiveness of Information-Decomposed Features}

\begin{figure*}[ht]
    \centering
    \includegraphics[scale=0.40]{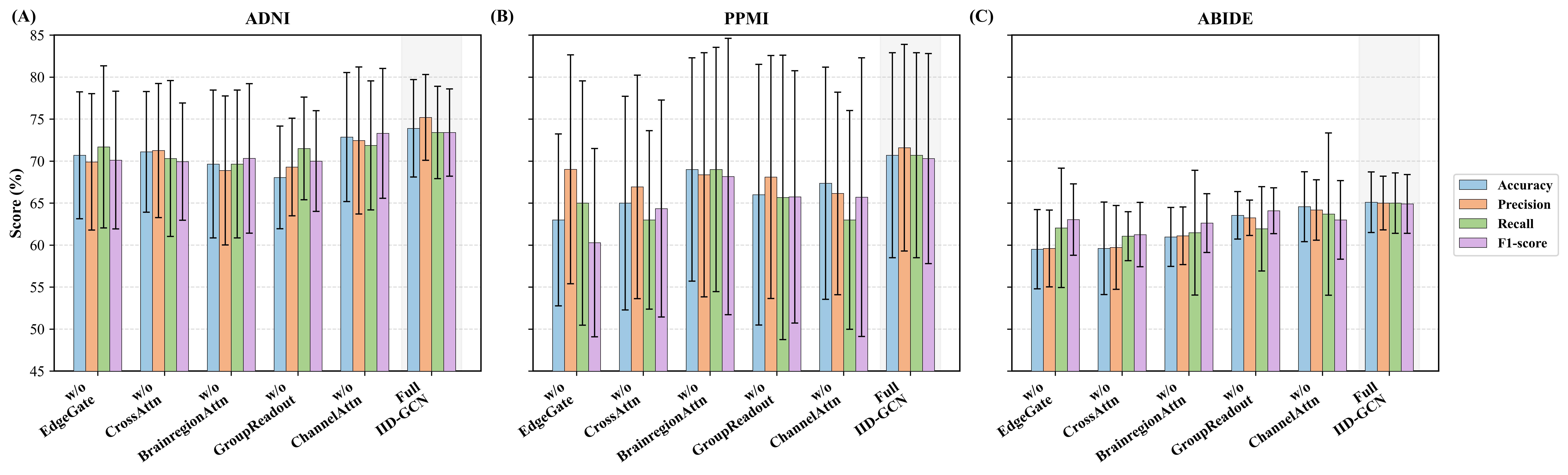}
   \caption{Ablation study of key modules in IID-GCN  on the ADNI, PPMI, and ABIDE datasets. The bars denote the mean classification scores, and the error bars denote the standard deviations across cross-validation folds.}
    \label{fig:module_ablation}
\end{figure*}

\begin{figure}[ht]
    \centering
    \includegraphics[scale=0.75]{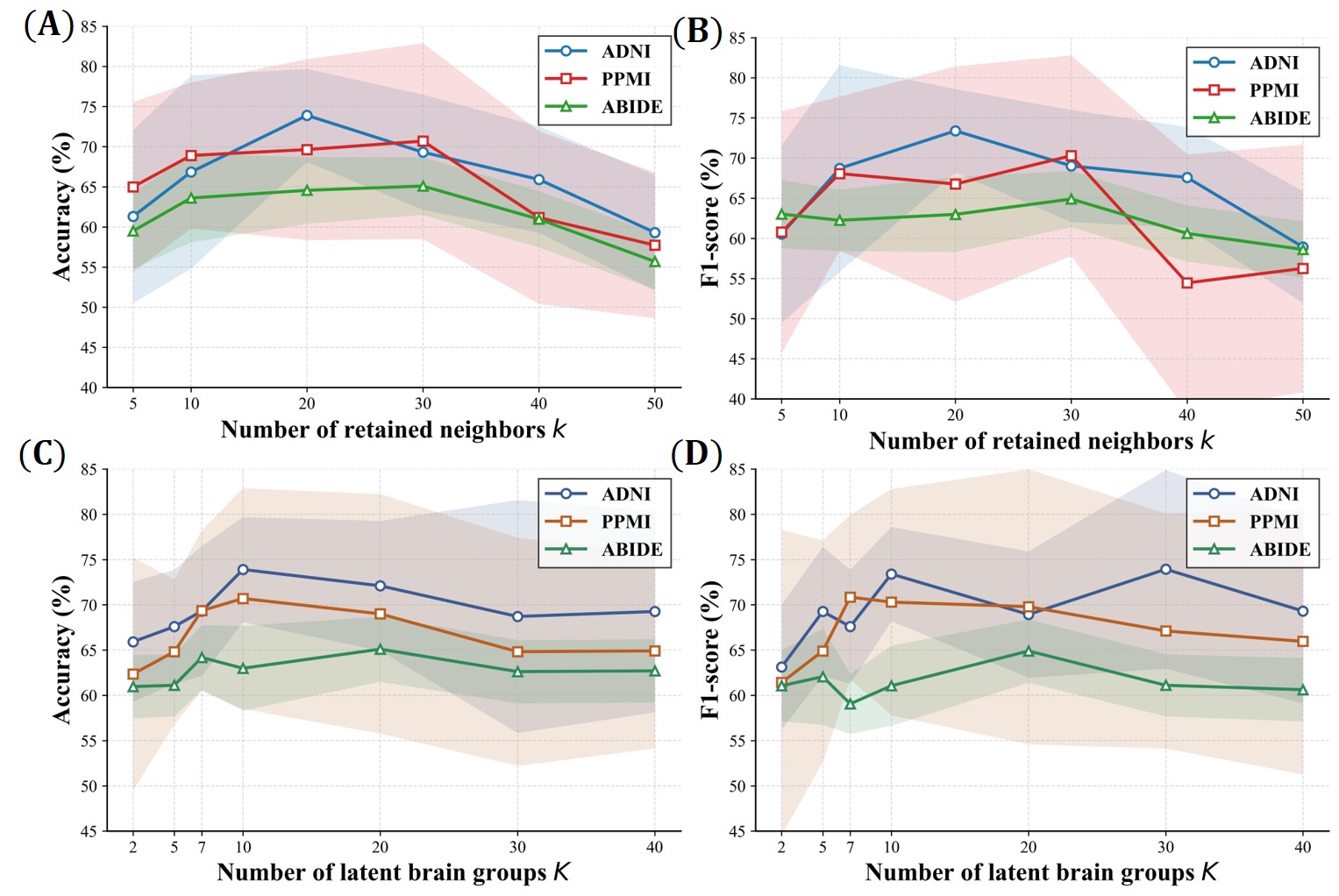}
   \caption{Hyperparameter analysis of IID-GCN on ADNI, PPMI, and ABIDE. We evaluate the effects of the retained neighbor number \(k\) and latent brain group number \(K\). The curves show the mean accuracy and F1-score across folds, and the shaded regions represent standard deviations.}
   \label{fig:hyperparameter_analysis}
\end{figure}

To assess the intrinsic discriminative value of information-decomposed features, we used SVM as a fixed classifier and separately evaluated Pearson correlation, redundancy, uniqueness, synergy, and their concatenation.  As shown in Table~\ref{tab:feature_ablation_svm}, the combined redundancy--uniqueness--synergy representation achieves the best overall performance across all datasets. Individual components show dataset-dependent strengths, with uniqueness performing well on ADNI and redundancy on PPMI, while the combined representation provides more balanced results on ABIDE. These results indicate that redundancy, uniqueness, and synergy provide additional discriminative information beyond traditional  functional connectivity.

\subsubsection{Effectiveness of Model Components}

To assess the contribution of each IID-GCN component, we conducted module-level ablations on all datasets by removing the residual edge gate, mutual cross-attention, ROI attention readout, soft node grouping readout, and channel attention module. As shown in Fig.~\ref{fig:module_ablation}, the full model consistently achieves the best overall performance, indicating that these modules provide complementary benefits. On ADNI, removing soft node grouping or ROI attention causes clear degradation, highlighting the importance of subnetwork- and ROI-level representations. On PPMI and ABIDE, removing the residual edge gate or mutual cross-attention leads to the largest performance drops, suggesting that adaptive edge recalibration and cross-information interaction are crucial for modeling information-decomposed brain graphs. These results support the integrated design of IID-GCN for robust rs-fMRI-based disease classification.

\subsection{Hyperparameter Analysis}

We further conducted a hyperparameter sensitivity analysis to evaluate the robustness of IID-GCN under different model configurations. Two key hyperparameters were examined: the number of retained neighbors $k$ for PED-based graph sparsification and the number of latent brain groups $K$. In each experiment, one parameter was varied while the others were fixed to the validation-selected default settings. The mean accuracy and F1-score across cross-validation folds are reported in Fig.~\ref{fig:hyperparameter_analysis}, with shaded regions denoting standard deviations.
\subsubsection{Analysis of graph sparsification parameter}
For the graph sparsification parameter $k$, performance generally improves from overly sparse graphs to a moderate range. Very small values, such as $k=5$, may discard informative disease-related connections, whereas overly large values can introduce weak or noisy information-theoretic edges. IID-GCN achieves stable performance around $k=20$ on ADNI and around $k=30$ on PPMI and ABIDE, suggesting that moderate graph sparsity is important for robust PED-based brain graph construction.
\subsubsection{Analysis of latent brain group number}
For the latent brain group number $K$, performance improves when $K$ increases from a very small value to a moderate range, indicating the benefit of modeling latent subnetwork-level organization. However, excessively large $K$ does not consistently improve performance and may introduce redundant or unstable group representations. The best or near-best results are typically obtained with a moderate number of latent groups.

\subsection{Visualization Analysis}

We saved subject-level cross-attention matrices from IID-GCN on the test set and averaged them within each diagnostic group to examine class-specific cross-information interactions.

\begin{figure*}[ht]
    \centering
    \includegraphics[scale=0.44]{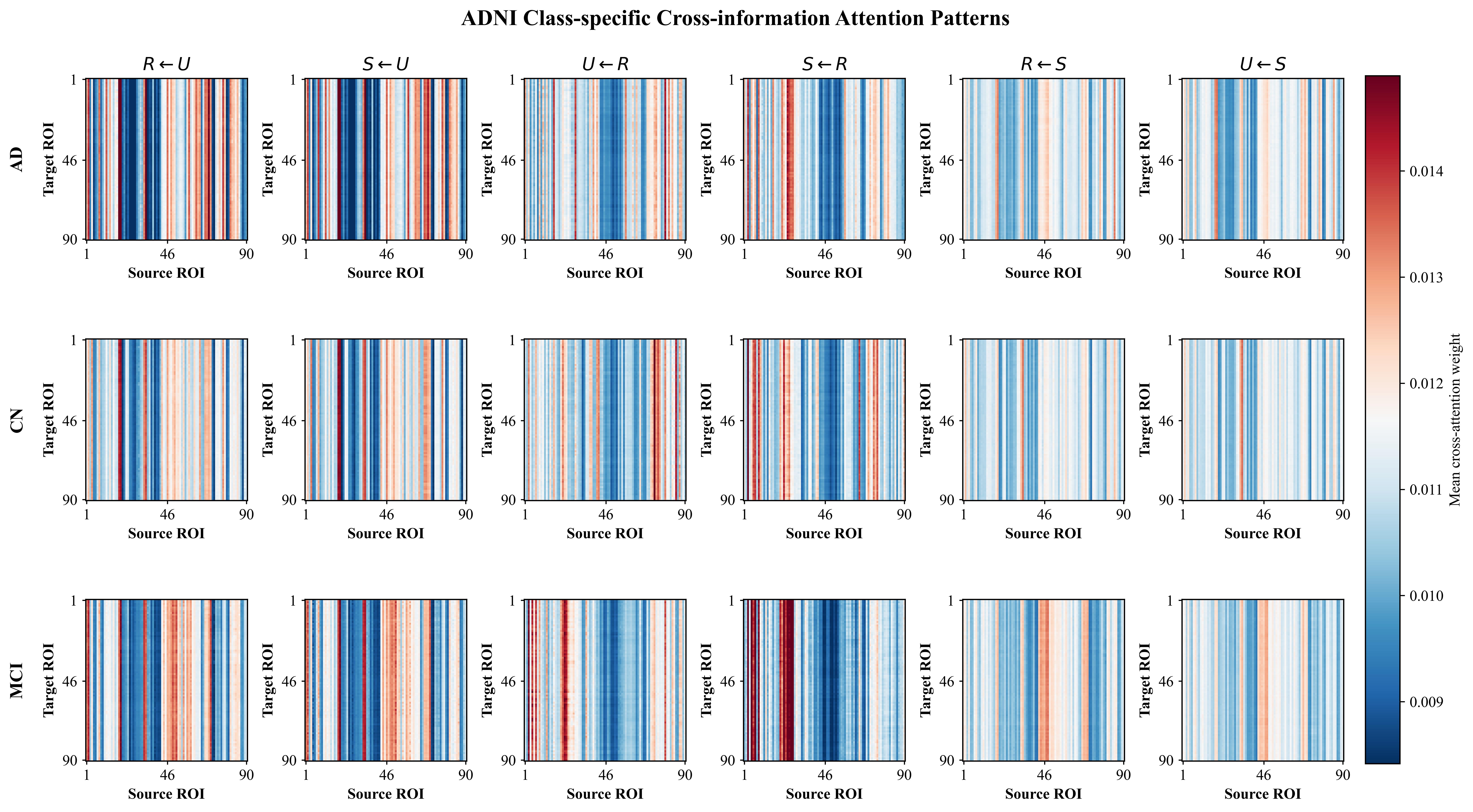}
   \caption{Class-specific cross-information interaction patterns learned by IID-GCN. Each heatmap shows the group-averaged cross-attention matrix..}
\label{fig:Cross-information interaction analysis}
\end{figure*}

As shown in Fig.~\ref{fig:Cross-information interaction analysis}, the attention maps are non-uniform and exhibit group-dependent patterns across AD, CN, and MCI. Pronounced vertical bands indicate that cross-channel information transfer is driven by specific source ROIs rather than uniformly distributed across regions. Compared with CN, AD and MCI show stronger deviations from the uniform attention baseline, particularly in directions involving redundancy and uniqueness, while MCI presents patterns partly distinct from both CN and AD. Similar non-uniform and diagnosis-dependent patterns are also observed on PPMI and ABIDE.

These results indicate that IID-GCN captures diagnosis-specific information exchange pathways, suggesting that brain disorders reshape functional information organization beyond traditional  connectivity strength.

\begin{figure*}[ht]
    \centering
    \includegraphics[scale=0.44]{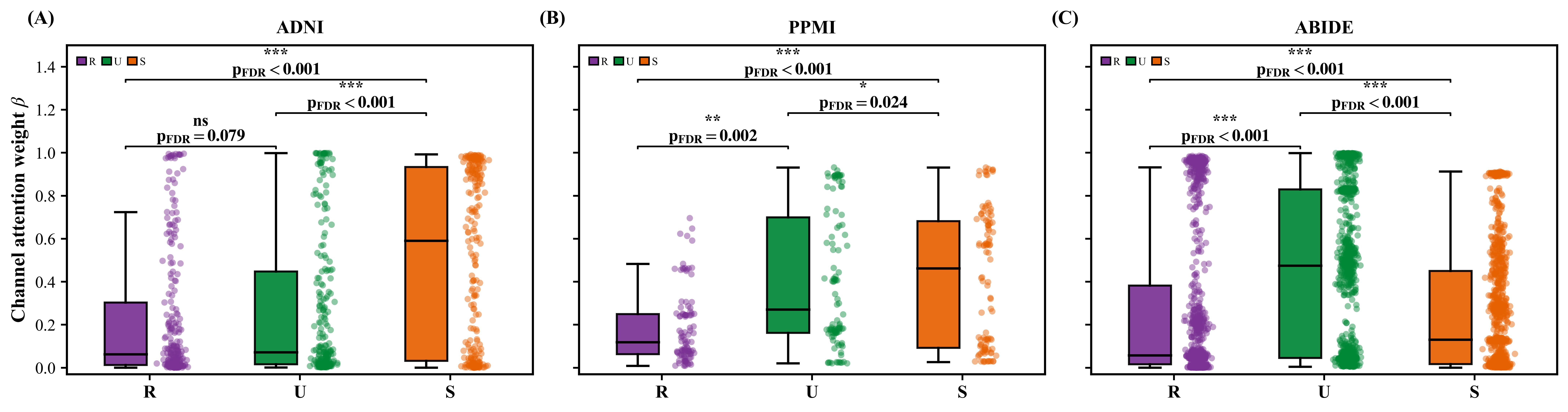}
   \caption{Information-channel contribution patterns learned by IID-GCN. Each box plot shows the overall distributions of channel-attention weights for redundancy, uniqueness, and synergy across all test subjects in each dataset. Higher attention weights indicate a stronger contribution of the corresponding information channel to the final graph-level representation. Pairwise statistical comparisons are marked above the box plots, with *, **, and *** denoting \(p<0.05\), \(p<0.01\), and \(p<0.001\), respectively.
   }
\label{Information-channel contribution analysis}
\end{figure*}

\begin{figure*}[ht]
    \centering
    \includegraphics[scale=0.54]{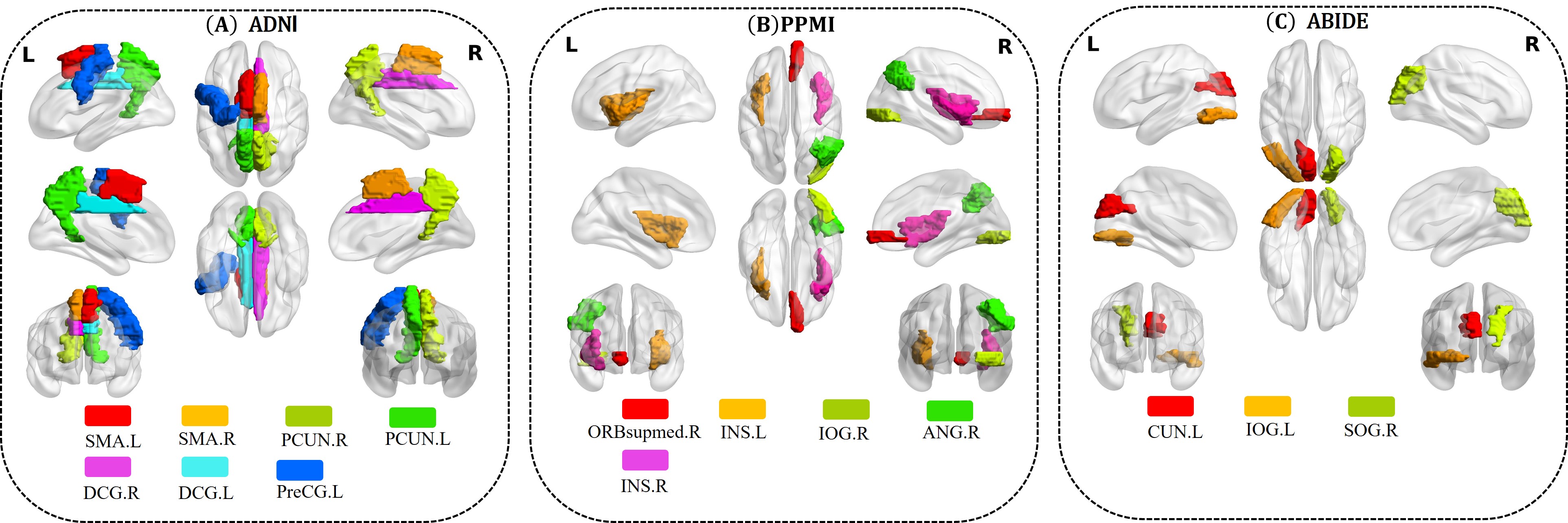}
   \caption{
Visualization of statistically selected important ROIs identified by the proposed IID-GCN. Panels show ROI importance maps across different datasets, computed from the dominant information channel indicated by channel-attention analysis: synergy for ADNI and PPMI, and uniqueness for ABIDE. Channel-specific attention scores were tested against a uniform baseline using permutation testing, followed by Benjamini--Hochberg FDR correction and bootstrap stability selection. Colored regions denote significant and stable model-attended ROIs, while the translucent gray surface provides anatomical context. L and R indicate the left and right hemispheres.
}
\label{fig:important_rois}
\end{figure*}

\subsection{Interpretability Analysis}
\begin{figure*}[ht]
    \centering
    \includegraphics[scale=0.38]{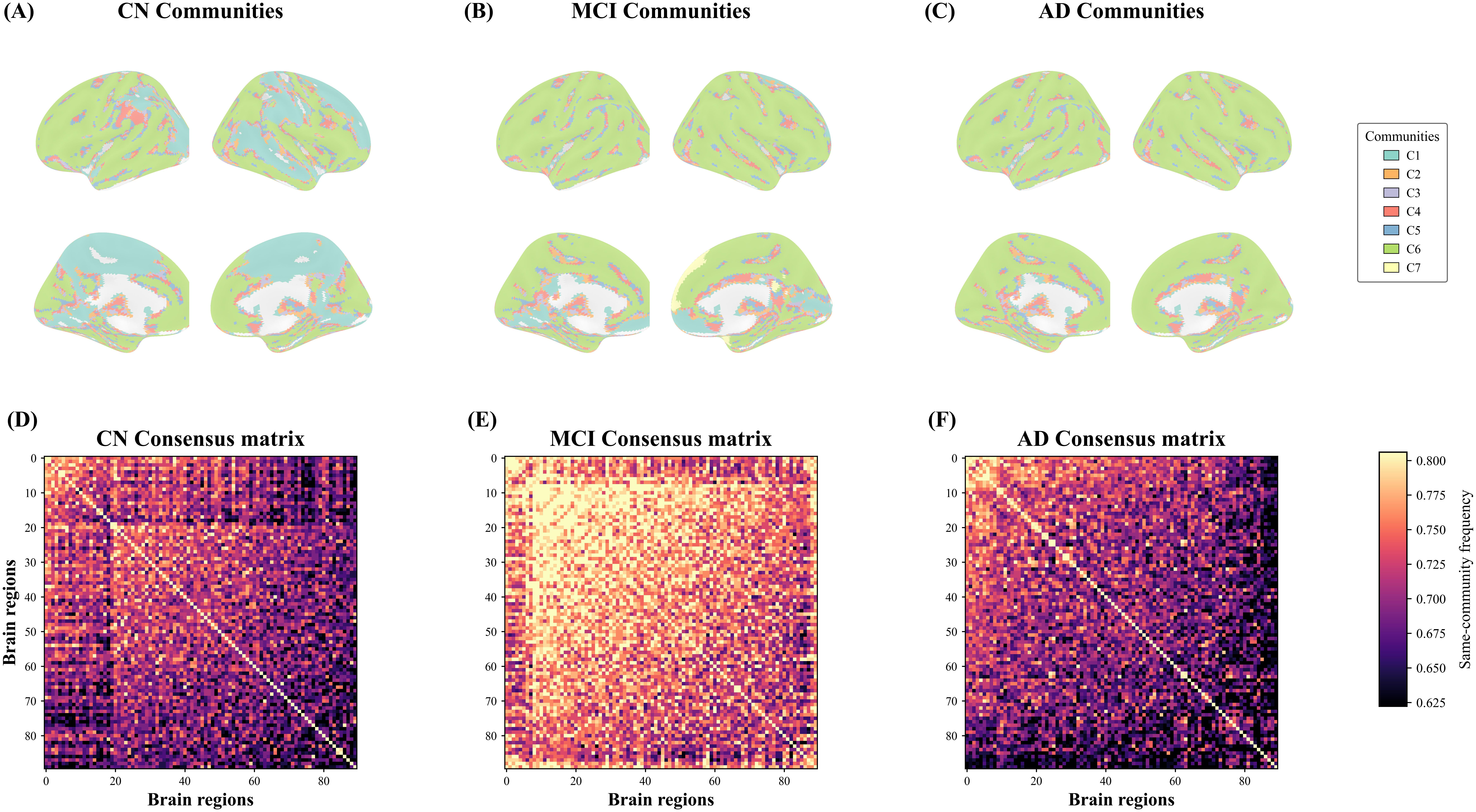}
   \caption{Class-specific synergy-channel subnetworks and consensus matrices.
Top row: surface visualization of class-specific community organization in the synergy channel for AD, CN, and MCI, where each cortical parcel is colored according to its dominant community assignment.
Bottom row: corresponding class-specific consensus matrices, where each entry represents the frequency with which a pair of ROIs is assigned to the same community across subjects.
}
\label{fig:subnetwork}
\end{figure*}
   

\subsubsection{Information-channel contribution analysis}
The result in Fig.~\ref{Information-channel contribution analysis} shows that IID-GCN learns distinct channel-attention patterns across datasets. On ADNI, the synergy channel receives the highest attention weight, significantly exceeding both redundancy and uniqueness, suggesting that jointly emergent information is particularly informative for Alzheimer's disease spectrum classification. On PPMI, synergy also shows the strongest contribution, while redundancy contributes the least, indicating the importance of synergistic information patterns for Parkinson’s disease analysis. In contrast, on ABIDE, the uniqueness channel dominates and is significantly higher than redundancy and synergy, suggesting that region-specific information plays a more important role in autism spectrum disorder classification.

Taken together, these results show that redundancy, uniqueness, and synergy contribute differently across disorders. The dominant information channel is disease-dependent, supporting the need for adaptive channel-attentive fusion rather than treating all information components equally.

\subsubsection{Brain-region importance analysis}
As shown in Fig.~\ref{fig:important_rois}, IID-GCN identifies dataset-specific important ROIs from the dominant information channels. On ADNI, significant ROIs are mainly distributed in motor, precuneus, cingulate, and precentral regions, suggesting altered synergy-related information organization in Alzheimer's disease spectrum classification. On PPMI, the selected ROIs include insular, angular, occipital, and orbitofrontal-related regions, indicating that synergy-driven information patterns contribute to Parkinson's disease discrimination. On ABIDE, important ROIs are mainly located in occipital regions, including the cuneus, inferior occipital gyrus, and superior occipital gyrus, suggesting a stronger role of uniqueness-related visual-network information in autism spectrum disorder classification.

Overall, the selected ROIs are anatomically meaningful and differ across datasets, indicating that IID-GCN captures disorder-specific information-channel patterns rather than relying on a fixed set of brain regions.

\subsubsection{Class-Specific Latent Subnetwork Analysis}
To interpret the latent subnetworks learned by soft node grouping, we visualized synergy-channel subnetworks and consensus matrices on ADNI \cite{varley2023partial}. For each subject, the soft assignment matrix was converted into ROI group labels by assigning each ROI to the group with the highest probability. We constructed a co-assignment matrix indicating whether each ROI pair belonged to the same group, and averaged these matrices within each diagnostic group to obtain consensus matrices.

As shown in Fig.~\ref{fig:subnetwork}, CN, MCI and AD show distinct synergy-channel subnetwork organizations. The consensus matrices exhibit different block-like patterns, indicating diagnosis-dependent latent subnetwork structures. These results suggest that IID-GCN captures disease-stage-related changes in synergy-based functional information organization.

\section{Conclusion}
\label{Conclusion}
In this study, we proposed IID-GCN, a novel graph convolutional network for rs-fMRI-based brain disease diagnosis. By decomposing pairwise functional interactions into redundancy, uniqueness, and synergy graphs, IID-GCN captures complementary information beyond traditional  correlation-based functional connectivity. The model jointly learns these information-specific graphs through multiple complementary modules for graph encoding. Experiments on ADNI, PPMI, and ABIDE demonstrate competitive performance in brain disorder diagnosis, while ablation and interpretability analyses confirm the value of the decomposed information graphs. The results suggest that brain disorders may affect not only connectivity strength, but also the organization of redundancy, uniqueness, and synergy across disease-relevant regions, supporting the potential of information-decomposed brain graphs as an interpretable representation.

\bibliographystyle{IEEEtran}
\bibliography{reference}

@article{fornito2015connectomics,
  title={The connectomics of brain disorders},
  author={Fornito, Alex and Zalesky, Andrew and Breakspear, Michael},
  journal={Nature reviews neuroscience},
  volume={16},
  number={3},
  pages={159--172},
  year={2015},
  publisher={Nature Publishing Group UK London}
}

@article{fox2007spontaneous,
  title={Spontaneous fluctuations in brain activity observed with functional magnetic resonance imaging},
  author={Fox, Michael D and Raichle, Marcus E},
  journal={Nature reviews neuroscience},
  volume={8},
  number={9},
  pages={700--711},
  year={2007},
  publisher={Nature Publishing Group UK London}
}

@article{van2010exploring,
  title={Exploring the brain network: a review on resting-state fMRI functional connectivity},
  author={Van Den Heuvel, Martijn P and Pol, Hilleke E Hulshoff},
  journal={European neuropsychopharmacology},
  volume={20},
  number={8},
  pages={519--534},
  year={2010},
  publisher={Elsevier}
}

@article{teng2023brain,
  title={Brain disease research based on functional magnetic resonance imaging data and machine learning: a review},
  author={Teng, Jing and Mi, Chunlin and Shi, Jian and Li, Na},
  journal={Frontiers in Neuroscience},
  volume={17},
  pages={1227491},
  year={2023},
  publisher={Frontiers Media SA}
}

@article{yin2022deep,
  title={Deep learning for brain disorder diagnosis based on fMRI images},
  author={Yin, Wutao and Li, Longhai and Wu, Fang-Xiang},
  journal={Neurocomputing},
  volume={469},
  pages={332--345},
  year={2022},
  publisher={Elsevier}
}

@article{hlinka2011functional,
  title={Functional connectivity in resting-state fMRI: is linear correlation sufficient?},
  author={Hlinka, Jaroslav and Palu{\v{s}}, Milan and Vejmelka, Martin and Mantini, Dante and Corbetta, Maurizio},
  journal={NeuroImage},
  volume={54},
  number={3},
  pages={2218--2225},
  year={2011},
  publisher={Elsevier}
}

@article{hartman2011role,
  title={The role of nonlinearity in computing graph-theoretical properties of resting-state functional magnetic resonance imaging brain networks},
  author={Hartman, D and Hlinka, J and Palu{\v{s}}, M and Mantini, Dante and Corbetta, Maurizio},
  journal={Chaos: An Interdisciplinary Journal of Nonlinear Science},
  volume={21},
  number={1},
  year={2011},
  publisher={AIP Publishing}
}

@article{wibral2017partial,
  title={Partial information decomposition as a unified approach to the specification of neural goal functions},
  author={Wibral, Michael and Priesemann, Viola and Kay, Jim W and Lizier, Joseph T and Phillips, William A},
  journal={Brain and cognition},
  volume={112},
  pages={25--38},
  year={2017},
  publisher={Elsevier}
}

@article{luppi2024synergistic,
  title={A synergistic workspace for human consciousness revealed by integrated information decomposition},
  author={Luppi, Andrea I and Mediano, Pedro AM and Rosas, Fernando E and Allanson, Judith and Pickard, John and Carhart-Harris, Robin L and Williams, Guy B and Craig, Michael M and Finoia, Paola and Owen, Adrian M and others},
  journal={Elife},
  volume={12},
  pages={RP88173},
  year={2024},
  publisher={eLife Sciences Publications, Ltd}
}

@article{varley2023partial,
  title={Partial entropy decomposition reveals higher-order information structures in human brain activity},
  author={Varley, Thomas F and Pope, Maria and Maria Grazia and Joshua and Sporns, Olaf},
  journal={Proceedings of the National Academy of Sciences},
  volume={120},
  number={30},
  pages={e2300888120},
  year={2023},
  publisher={National Academy of Sciences}
}

@article{bessadok2022graph,
  title={Graph neural networks in network neuroscience},
  author={Bessadok, Alaa and Mahjoub, Mohamed Ali and Rekik, Islem},
  journal={IEEE Transactions on Pattern Analysis and Machine Intelligence},
  volume={45},
  number={5},
  pages={5833--5848},
  year={2022},
  publisher={IEEE}
}

@inproceedings{yan2019groupinn,
  title={Groupinn: Grouping-based interpretable neural network for classification of limited, noisy brain data},
  author={Yan, Yujun and Zhu, Jiong and Duda, Marlena and Solarz, Eric and Sripada, Chandra and Koutra, Danai},
  booktitle={Proceedings of the 25th ACM SIGKDD international conference on knowledge discovery \& data mining},
  pages={772--782},
  year={2019}
}

@article{mohammadi2024graph,
  title={Graph neural networks in brain connectivity studies: Methods, challenges, and future directions},
  author={Mohammadi, Hamed and Karwowski, Waldemar},
  journal={Brain Sciences},
  volume={15},
  number={1},
  pages={17},
  year={2024},
  publisher={MDPI}
}

@article{luo2024graph,
  title={Graph neural networks for brain graph learning: A survey},
  author={Luo, Xuexiong and Wu, Jia and Yang, Jian and Xue, Shan and Beheshti, Amin and Sheng, Quan Z and McAlpine, David and Sowman, Paul and Giral, Alexis and Yu, Philip S},
  journal={arXiv preprint arXiv:2406.02594},
  year={2024}
}

@article{alarjani2025brain,
  title={Brain functional connectivity analysis of fMRI-based Alzheimer's disease data},
  author={Alarjani, Maitha S. and Almarri, Badar A.},
  journal={Frontiers in Medicine},
  volume={12},
  pages={1540297},
  year={2025},
  doi={10.3389/fmed.2025.1540297}
}

@article{tessadori2025linking,
  title={Linking dynamic connectivity states to cognitive decline and anatomical changes in Alzheimer's disease},
  author={Tessadori, Jacopo and Boscolo Galazzo, Ilaria and Storti, Silvia F. and Pini, Lorenzo and Brusini, Lorenza and Cruciani, Federica and Sona, Diego and Menegaz, Gloria and Murino, Vittorio},
  journal={NeuroImage},
  volume={320},
  pages={121448},
  year={2025},
  doi={10.1016/j.neuroimage.2025.121448}
}

@article{karim2025functional,
  title={Functional connectivity signatures in fMRI-derived connectome for the diagnosis of autism spectrum disorder},
  author={Karim, S. M. Shayez and Rathore, R. S.},
  journal={Brain Organoid and Systems Neuroscience Journal},
  volume={3},
  pages={170--179},
  year={2025},
  doi={10.1016/j.bosn.2025.06.004}
}

@article{tang2025gnnfmri,
  title={Graph neural networks for fMRI functional brain networks: A survey},
  author={Tang, Jingye and Zhu, Tianqing and Zhou, Wanlei and Zhao, Wei},
  journal={Neural Networks},
  volume={194},
  pages={108137},
  year={2025},
  doi={10.1016/j.neunet.2025.108137}
}

@article{ali2025gnnad,
  title={Graph neural networks in Alzheimer's disease diagnosis: A review of unimodal and multimodal advances},
  author={Ali, Shahzad and Piana, Michele and Pardini, Matteo and Garbarino, Sara},
  journal={Frontiers in Neuroscience},
  volume={19},
  pages={1623141},
  year={2025},
  doi={10.3389/fnins.2025.1623141}
}

@article{sun2025admgcn,
  title={ADMGCN: Graph convolutional network for Alzheimer's disease diagnosis with a meta-learning paradigm},
  author={Sun, Xiaowen and others},
  journal={Bioinformatics},
  volume={41},
  number={12},
  pages={btaf580},
  year={2025},
  doi={10.1093/bioinformatics/btaf580}
}

@article{li2021braingnn,
  title={BrainGNN: Interpretable brain graph neural network for fMRI analysis},
  author={Li, Xiaoxiao and Zhou, Yuan and Dvornek, Nicha C. and Zhang, Muhan and Gao, Siyuan and Zhuang, Juntang and Scheinost, Dustin and Staib, Lawrence H. and Ventola, Pamela and Duncan, James S.},
  journal={Medical Image Analysis},
  volume={74},
  pages={102233},
  year={2021}
}

@article{ji2022fchat,
  title={FC-HAT: Hypergraph attention network for functional brain network classification},
  author={Ji, Junzhong and Ren, Yating and Lei, Minglong},
  journal={Information Sciences},
  volume={608},
  pages={1301--1316},
  year={2022}
}

@article{zhang2023local,
  title={Classification of brain disorders in rs-fMRI via local-to-global graph neural networks},
  author={Zhang, Hao and Song, Ran and Wang, Liping and Zhang, Lin and Wang, Dawei and Wang, Cong and Zhang, Wei},
  journal={IEEE Transactions on Medical Imaging},
  volume={42},
  number={2},
  pages={444--455},
  year={2023}
}

@article{wang2023ucgl,
  title={Unsupervised contrastive graph learning for resting-state functional MRI analysis and brain disorder detection},
  author={Wang, X. and Chu, Y. and Wang, Q. and Cao, L. and Qiao, L. and Zhang, L. and Liu, M.},
  journal={Human Brain Mapping},
  volume={44},
  number={17},
  pages={5672--5692},
  year={2023}
}

@article{ince2017partial,
  title={The Partial Entropy Decomposition: Decomposing multivariate entropy and mutual information via pointwise common surprisal},
  author={Ince, Robin AA},
  journal={arXiv preprint arXiv:1702.01591},
  year={2017}
}

@article{dadi2019benchmarking,
  title={Benchmarking functional connectome-based predictive models for resting-state fMRI},
  author={Dadi, Kamalaker and Rahim, Mehdi and Abraham, Alexandre and Chyzhyk, Darya and Milham, Michael and Thirion, Bertrand and Varoquaux, Ga{\"e}l and Alzheimer's Disease Neuroimaging Initiative and others},
  journal={NeuroImage},
  volume={192},
  pages={115--134},
  year={2019},
  publisher={Elsevier}
}

@article{craddock2013neuro,
  title={The neuro bureau preprocessing initiative: open sharing of preprocessed neuroimaging data and derivatives},
  author={Craddock, Cameron and Benhajali, Yassine and Chu, Carlton and Chouinard, Francois and Evans, Alan and Jakab, Andr{\'a}s and Khundrakpam, Budhachandra Singh and Lewis, John David and Li, Qingyang and Milham, Michael and others},
  journal={Frontiers in Neuroinformatics},
  volume={7},
  number={27},
  pages={5},
  year={2013}
}

@article{xu2023data,
  title={Data-driven network neuroscience: On data collection and benchmark},
  author={Xu, Jiaxing and Yang, Yunhan and Huang, David and Gururajapathy, Sophi Shilpa and Ke, Yiping and Qiao, Miao and Wang, Alan and Kumar, Haribalan and McGeown, Josh and Kwon, Eryn},
  journal={Advances in Neural Information Processing Systems},
  volume={36},
  pages={21841--21856},
  year={2023}
}

@article{esteban2019fmriprep,
  title={fMRIPrep: a robust preprocessing pipeline for functional MRI},
  author={Esteban, Oscar and Markiewicz, Christopher J and Blair, Ross W and Moodie, Craig A and Isik, A Ilkay and Erramuzpe, Asier and Kent, James D and Goncalves, Mathias and DuPre, Elizabeth and Snyder, Madeleine and others},
  journal={Nature methods},
  volume={16},
  number={1},
  pages={111--116},
  year={2019},
  publisher={Nature Publishing Group US New York}
}

@article{tzourio2002automated,
  title={Automated anatomical labeling of activations in SPM using a macroscopic anatomical parcellation of the MNI MRI single-subject brain},
  author={Tzourio-Mazoyer, Nathalie and Landeau, Brigitte and Papathanassiou, Dimitri and Crivello, Fabrice and Etard, Octave and Delcroix, Nicolas and Mazoyer, Bernard and Joliot, Marc},
  journal={Neuroimage},
  volume={15},
  number={1},
  pages={273--289},
  year={2002},
  publisher={Elsevier}
}

@article{pedregosa2011scikit,
  title={Scikit-learn: Machine learning in Python},
  author={Pedregosa, Fabian and Varoquaux, Ga{\"e}l and Gramfort, Alexandre and Michel, Vincent and Thirion, Bertrand and Grisel, Olivier and Blondel, Mathieu and Prettenhofer, Peter and Weiss, Ron and Dubourg, Vincent and others},
  journal={the Journal of machine Learning research},
  volume={12},
  pages={2825--2830},
  year={2011},
  publisher={JMLR. org}
}

@article{kipf2016semi,
  title={Semi-supervised classification with graph convolutional networks},
  author={Kipf, Thomas N and Welling, Max},
  journal={arXiv preprint arXiv:1609.02907},
  year={2016}
}

@article{hamilton2017inductive,
  title={Inductive representation learning on large graphs},
  author={Hamilton, Will and Ying, Zhitao and Leskovec, Jure},
  journal={Advances in neural information processing systems},
  volume={30},
  year={2017}
}

@article{velivckovic2017graph,
  title={Graph attention networks},
  author={Veli{\v{c}}kovi{\'c}, Petar and Cucurull, Guillem and Casanova, Arantxa and Romero, Adriana and Lio, Pietro and Bengio, Yoshua},
  journal={arXiv preprint arXiv:1710.10903},
  year={2017}
}

@inproceedings{kan2022fbnetgen,
  title={Fbnetgen: Task-aware gnn-based fmri analysis via functional brain network generation},
  author={Kan, Xuan and Cui, Hejie and Lukemire, Joshua and Guo, Ying and Yang, Carl},
  booktitle={International conference on medical imaging with deep learning},
  pages={618--637},
  year={2022},
  organization={PMLR}
}

@article{zheng2024bpi,
  title={BPI-GNN: Interpretable brain network-based psychiatric diagnosis and subtyping},
  author={Zheng, Kaizhong and Yu, Shujian and Chen, Liangjun and Dang, Lujuan and Chen, Badong},
  journal={NeuroImage},
  volume={292},
  pages={120594},
  year={2024},
  publisher={Elsevier}
}

@article{bian2023adversarially,
  title={Adversarially trained persistent homology based graph convolutional network for disease identification using brain connectivity},
  author={Bian, Chenyuan and Xia, Nan and Xie, Anmu and Cong, Shan and Dong, Qian},
  journal={IEEE Transactions on Medical Imaging},
  volume={43},
  number={1},
  pages={503--516},
  year={2023},
  publisher={IEEE}
}

@article{zeng2025adaptive,
  title={Adaptive multi-scale dynamic graph representation learning with overlapping community-awareness for ASD classification},
  author={Zeng, Wenwen and Yin, Feiyu and Song, Pengfei and Wu, Yonghuang and Zhao, Chengqian and Wu, Guoqing and Yu, Jinhua},
  journal={IEEE Journal of Biomedical and Health Informatics},
  volume={29},
  number={12},
  pages={8711--8718},
  year={2025},
  publisher={IEEE}
}

@article{kingma2014adam,
  title={Adam: A method for stochastic optimization},
  author={Kingma, Diederik P and Ba, Jimmy},
  journal={arXiv preprint arXiv:1412.6980},
  year={2014}
}

@article{zhao2026hoi,
  title={HOI-brain: A novel multi-channel transformers framework for brain disorder diagnosis by accurately extracting signed higher-order interactions from fMRI data},
  author={Zhao, Dengyi and Zhou, Zhiheng and Yan, Guiying and Yu, Dongxiao and Qi, Xingqin},
  journal={Medical Image Analysis},
  pages={104009},
  year={2026},
  publisher={Elsevier}
}

@article{zhou2024novel,
  title={A novel graph neural network method for Alzheimer’s disease classification},
  author={Zhou, Zhiheng and Wang, Qi and An, Xiaoyu and Chen, Siwei and Sun, Yongan and Wang, Guanghui and Yan, Guiying},
  journal={Computers in Biology and Medicine},
  volume={180},
  pages={108869},
  year={2024},
  publisher={Elsevier}
}

@article{zhao2026extracting,
  title={Extracting interpretable higher-order topological features across multiple scales from fMRI for Alzheimer’s disease classification},
  author={Zhao, Dengyi and Li, Shanyong and Wang, Yunping and Wang, Chenfei and Zhou, Zhiheng and Yan, Guiying and Qi, Xingqin},
  journal={Biomedical Signal Processing and Control},
  volume={116},
  pages={109621},
  year={2026},
  publisher={Elsevier}
}

@article{wang2026classification,
  title={Classification of Alzheimer's Disease by Modeling Brain Networks as Signed Networks under Deep Learning Frameworks},
  author={Wang, Chenfei and Wang, Yunping and Xue, Qinghan and Zhou, Zhiheng and Yan, Guiying and Qi, Xingqin},
  journal={IEEE Transactions on Computational Biology and Bioinformatics},
  year={2026},
  publisher={IEEE}
}

\end{document}